\documentclass[twocolumn,showpacs,preprintnumbers,amsmath,amssymb]{revtex4}

\usepackage{graphicx}
\usepackage{dcolumn}
\usepackage{bm}
\newcommand{\etal}{\emph{et al.}}
\newcommand{\be}{\begin{equation}}
\newcommand{\ee}{\end{equation}}
\newcommand{\bfig}{\begin{figure}}
\newcommand{\efig}{\end{figure}}
\newcommand{\incl}{\includegraphics}

\begin{document}      

\title{Speeding up a single-molecule DNA device with a simple catalyst} 
\author{Yufang Wang, Y. Zhang$^\dagger$, and N. P. Ong}      
\affiliation{Department of Physics, 
Princeton University, Princeton, New Jersey 08544, U.S.A.}

\date{\today}      

\begin{abstract}

Recently, several groups have designed and synthesized single-molecule devices 
based on DNA that can switch between different
configurations in response to sequential addition of fuel DNA strands.  
There is considerable interest in improving the speed of these `nanomotors'.  
One approach is the use of rationally designed DNA catalysts to 
promote hybridization of complementary oligonucleotides.  A particularly simple and robust DNA device
reported by Li and Tan is comprised of a single-strand 17-base oligomer that folds into a chair-like 
quadruplex structure.  We have identified the key rate-limiting barrier in this device as 
the tendency for one of the fuel strands $B$ to fold into the quadruplex configuration of the device strand.  
This seriously impedes the restoration reaction.  We have designed a catalytic strand to inhibit the 
folding of $B$, and shown that the catalyst speeds up the restoration reaction by roughly a factor of 2.  
The catalyst remains effective even after repeated cycling.  
\end{abstract}
\maketitle                   

\section{Introduction}\label{intro}
The operation of nanoscale devices based on 
designed DNA strands capable of switching between configurations in response to
the addition of DNA fuel molecules in a cyclic 
manner has been demonstrated by several groups~\cite{Mao,Yurke1,Yurke2,Tan,Yurke3,Yan}.  The different 
configurations and the sequence recognition properties of DNA have 
been utilized to induce mechanical motion on a molecular scale.  
Mao \etal~\cite{Mao} exploited the transition between the B and Z forms 
of DNA molecules to induce atomic displacement of 20 to 60 \AA.  
The construction of nanomotors based on DNA molecules and 
branch migration has been pioneered by Yurke and his collaborators.  These molecular devices,
dubbed ``nanomotors'', exploit configuration changes induced by hybridization reactions.

Yurke and collaborators have designed DNA tweezers~\cite{Yurke1}, 
a DNA actuator~\cite{Yurke2}, as well as a device that is switchable between three 
mechanical states~\cite{Yurke3}.  The transition between different states may be 
induced by a change in experimental conditions or by the addition 
of a ``DNA fuel" that drives the cycle.  Yan \etal~\cite{Yan} have also 
reported a robust sequence-dependent rotary DNA device that flips 
between two well-defined stable configurations.  Progress in the design of single-molecule 
DNA devices with low molecular weight seems encouraging for future applications in 
nanotechnoglogy.  A well-designed single-molecule DNA device should 
display a robust transition between the straightened state and a folded 
state, examples of which are the G-quadruplexes~\cite{Williamson,Alberti},  
and the hybridized hairpin structure~\cite{Feng}.  

\section{Catalytic control of the DNA device}\label{catalyst}
Recently, Li and Tan~\cite{Tan} introduced a single-molecule 
device based on a short DNA oligomer.  In their design,
the `motor' strand $M$ is a 17-base single strand (5'-TGG TTG GTG TGG TTG GT-3')
that folds up into a ``chair" quadruplex structure in the presence of potassium 
ions (see Fig. \ref{strands}a).  The cycle is driven by the introduction of 
the 27-base fuel strand 
$A$ (5'-GTA GTC CGC GAC CAA CCA CAC CAA CCA-3'), 
which consists of a 17-base segment that is the Watson-Crick complement 
of $M$, and a 10-base overhang.  Because the persistence length of double-stranded 
DNA is about 100 base pairs~\cite{Smith}, $M$ straightens 
out into a rigid double strand on hybridizing with $A$
(Fig. \ref{strands}b).  It may then be restored to its initial chair configuration by the introduction 
of the restoration fuel $B$ (5'-TGG TTG GTG TGG TTG GTC GCG GAC TAC-3'), 
which is the complement of $A$.  The strand $B$ first attaches itself to $A$ via the 
10-base overhang.  Then it competes with $M$ for binding with $A$ through 
the process of branch migration~\cite{Green}.  When $M$ falls off from $A$, 
it returns to the folded chair structure, completing the cycle.  The complex $AB$
(the waste product) does not participate further in the cycle.
By successively adding $A$ and then $B$, the motor strand $M$ can be repeatedly 
cycled between its straightened and folded configurations.

As shown in Fig. \ref{strands}a, the strand $M$ is labeled at its 5' and 3' 
ends with the fluorophore 6-FAM 
(5'-fluorescein phosphoramidite) and the quencher DABCYL 
[4-(4'-dimethylaminophenylazo) benzoic acid], respectively.  
At room temperature, the excitation and emission peak wavelength of the 6-FAM dye 
are 492 nm and 518 nm, respectively. The molecule DABCYL is a wide-spectrum quencher 
that quenches efficiently the fluorescence of 6-FAM.
Because fluorescence is unaffected (strongly quenched) 
when the ends are far apart (in close proximity), the fluorescence 
intensity is a reliable indicator of the configuration change in $M$.

In the device of Li and Tan~\cite{Tan}, the fuel 
strand $B$ has a segment that is identical with $M$ (by complementarity).  
Hence it shares the latter's tendency to fold.  Folding of a fraction of 
$B$'s explains why the restoration reaction is significantly slower than 
the straightening reaction.  The method of mutated fuel was 
developed by introducing mismatches between the motor strand and 
straightening fuel~\cite{Alberti,Feng}.  This partially solves the problem, but at the 
cost of slowing the straightening reaction.  In a recent experiment, Turberfield \etal~\cite{Turberfield}
used loop complexes to inhibit the hybridization of complementary oligonucleotides. 
They found that rationally designed DNA catalysts are effective in promoting the hybridization.


Here we report a catalyst that specifically applies to the DNA 
device of Li and Tan~\cite{Tan}.  The catalyst, which works by binding to the 
fuel strand $B$ in order to inhibit its folding, consists of a short strand $C$ (5'-CGC GAC CA-3') 
that hybridizes to 2 of the 8 guanines in $B$ that form the quadruplex.  
In addition, $C$ hybridizes with other nearby bases in the 
overhang of $B$ to stabilize the binding.  With 1 M monovalent cation, at the concentration
of 1 $\mu$M, the melting temperature of $BC$ is 45.6$^\mathrm{o}$C~\cite{Peyret}, 
which is higher than room temperature at which the device is operated.
(In the absence of the overhang $C$ hybridizes with only 4 bases in $M$.  As the melting
temperature of the complex $CM$ is below room temperature, $C$ does not form a stable
duplex with $M$.)  If the folding of $B$ is the reason for slow restoration, the catalyst $C$ should accelerate 
the reaction (Fig. \ref{strands}c).

\section{Experimental results}\label{expt}
All oligonucleotides used in the experiments were synthesized and purified by 
Integrated DNA Technologies, Inc. (IDT).  Stock solutions were prepared by 
resuspending the lyophilized oligonucleotides in sodium phosphate/sodium 
chloride (SPSC) buffer (pH 6.5, 50 mM $\rm Na_2HPO_4$, 1 M NaCl) at the 
concentration of 100 $\mu$M. The experiments were performed at 1 $\mu$M 
concentration at 20$^{\mathrm{o}}$C by diluting stock solution of $M$ 100 fold in 
KSPSC (pH 6.5, 40 mM KCl, 50 mM $\rm Na_2HPO_4$, 1 M NaCl) buffer.  
The stock solutions of strand $A$ and $B$ are added and mixed by rapid 
pipetting up and down for $\sim$10 s.

Time-based fluorescence intensity curves were taken on a custom-designed 
spectrofluorometer (Photon Technology International), using a Hellma 
105.254-QS cuvette.  The excitation wavelength was set at 492 nm and 
the emission wavelength at 518 nm.  The excitation light intensity was 
kept low so that during the experiment the fluorescence bleaching effect 
was negligible compared with the effect caused by configuration changes in $M$.
%

Figure \ref{decay} compares the time dependence of the fluorescence 
intensity with the catalyst present (Sample $a$)
or absent ($b$) as the device is cycled by successive addition 
of fuel $A$ and $B$.  In both traces, the straightening 
reaction on addition of $A$ is too rapid to be resolved.  By contrast, 
the restoration reaction following the addition of $B$ 
is slower.  Comparing the two curves, we observe that
the addition of $C$ does not affect the steep rise when $A$ is added
(within our resolution capability), but significantly accelerates the restoration reaction 
following the addition of $B$. 

The vertical jumps in both traces 
when $A$ or $B$ is added (at time $t =$ 60 and 120 s, respectively) is an artifact of our 
experimental procedure.  Right before $A$ or $B$ is added, the shutter of the spectrofluorometer 
is turned off for a short `blind' interval ($\Delta t \simeq$ 10 s) to protect the photo-sensors
from strong light.  During this interval, the solution is repeatedly pipetted to improve mixing, but data are not recorded.  
The relatively fast reaction time ($\sim 100$ s) in the Li-Tan device makes this `blind' 
interval much more noticeable than in motors with longer reaction times.  
However, the numerical fitting procedure (Appendix) allows us
to interpolate the data within $\Delta t$.  (During $\Delta t$, the highly inhomogeneous concentration 
of $B$ prior to mixing seems to accelerate the reaction greatly in some regions of the solution.  Our
simulations favor a longer effective $\Delta t$ than the 10 s when the spectrometer
is actually in the off state.)

The effect of varying the concentration of the catalyst $C$ on the restoration 
reaction has been investigated in detail in an ensemble of 24 samples (at 8 different concentrations).  
To minimize the experimental uncertainties, these samples were all prepared at the same time, 
kept wrapped in aluminum foil and stored at 4$^{\mathrm{o}}$C until the measurement.  
In Fig. \ref{modelfit}a, three curves $a$, $b$ and $c$ at successively higher concentration 
of $C$ are shown in linear scale.  Clearly, the rate of decrease is accelerated with increasing concentration of
$C$.  In each case, rather close fits are obtained using the 
model described below (solid lines).  The fits allow interpolation within the
blind interval just after $B$ is added.

Although numerical integration of the rate equations is necessary to describe the curves accurately, 
we note that at early times (semilog plot in Fig. \ref{modelfit}b), they are well-described 
by a single exponential term.  The straight-line behavior in early times determines an 
effective time constant $\tau$ which we have used to characterize the effect of increasing $C$ 
(at latter times the experimental curves deviate noticeably from simple exponential decay).  
The dependence of $\tau$ on the concentration 
of $C$ measured in the 24 samples is displayed in Fig. \ref{tau}, which is our central experimental finding.  
At low concentrations, the catalyst speeds up
the reaction rate dramatically but the improvement saturates at moderately high concentrations.  [We emphasize 
that the absence of data at the very early interval $\Delta t$ (`vertical jump' discussed above) does not 
pose a problem for the semi-log plot in Fig. \ref{modelfit}b, as it merely shifts the origin of $t$.  
Using the semi-log plot, the effective time constant $\tau$ is directly inferred from the raw data.  
Hence the variation of $\tau$ shown in Fig. \ref{tau} is independent of our model.]

The catalyst remains efficient under repeated cycling.  As shown in 
Fig. \ref{repeatcycles}, the restoration reaction with catalyst is reproducibly 
faster than that without catalyst.

\section{Analysis}\label{model}
The operation of this DNA nano-device is driven by the hybridization of Watson-Crick 
complementary sections of DNA.  In the absence of the catalyst, two processes 
are involved in the restoration reaction.  First, $B$ hybridizes with the single-stranded 
overhang section of $A$.  Following the opening of $B$, random branch migration along 
the motor-strand section of $MA$ occurs until, either the waste product $AB$ is formed, and $M$ 
returns to its folded state, or the branch point returns to its initial point and $B$ 
dissociates from the complex (or branch migration starts anew).

As observed in our experiment, the restoration is always much slower than the 
straightening.  In branch migration, the random walk process is associated with 
the temporary breaking of base pair bonds by thermal fluctuations.  
When uninhibited, a random walk over short segments (10-15 base pairs)
is expected to take about 15-20 ms~\cite{Yurke4}, which is much shorter than the
observed reaction time.  Hence something else is impeding restoration.  
An important factor here is the tendency of the unhybridized 
$B$ to collapse into its folded state.  We reason that the folding 
of $B$ strongly inhibits the initiation of branch migration right after the overhang sections of $B$ and
$A$ become hybridized (see Fig. \ref{strands}b).  Hence this is most likely the reason for
the relatively long restoration time of the cycle.

In the ``all-or-none" model of Morrison and Stols~\cite{Morrison}, 
the hybridization reactions of DNA is described by the equation
\be
S + \overline{S}\; \mathop{\rightleftharpoons}_{k_-}^{k_+} \;S\overline{S},
\label{SS}
\ee
where $S$ and $\overline{S}$ are a pair of complementary DNA strands, and $k_+$ and $k_-$ 
are the forward-reaction and off rate constants, respectively.

Our model is then described by the following reactions 
\begin{eqnarray}
B + C &\stackrel{k_0}{\longrightarrow}& BC\label{BC} \\
MA + B &\stackrel{k_{1}}{\longrightarrow}& MAB \label{MA}\\
MAB &\stackrel{k_{2}}{\longrightarrow}& M + AB \label{MAB}\\
MA + BC &\stackrel{k_{3}}{\longrightarrow}& MAB' + C \label{MABC}\\
MAB' &\stackrel{k_{4}}{\longrightarrow}& M + AB \label{M},
\end{eqnarray} 
where $k_0,\cdots, k_4$ are the corresponding forward reaction rate constants
(the subscript $+$ is suppressed hereafter).  In the Appendix, we show 
that the corresponding off reaction rate constants are negligible.

Equations \ref{BC}-\ref{M}, which describe 2 parallel 
processes for the production of the final products $M$ and $AB$ (`sinks') starting
with the products $MA$ and $B$ (`sources'), may be
represented by the simple diagram in Fig. \ref{rate}.  
The `slow' process (Eqs. \ref{MA} and \ref{MAB}) is independent of $C$, and 
involves the hybridization of $MA$ and $B$ to form $MAB$ which 
subsequently decays to $M$ and $AB$
(upper branch in Fig. \ref{rate}).  Its rate constants, $k_1$ and $k_2$, 
are readily obtained by fitting the measured decay curves in which $c_0$ = 0.  
(see Appendix).

The `fast' process, which is catalyzed by $C$, 
is initiated by the formation of the duplex $BC$ (Eqs. \ref{BC} , \ref{MABC} and \ref{M}; 
lower branch in Fig. \ref{rate}).
Hybridization of $BC$ and $MA$ produces the unstable 4-strand complex $MABC$ 
which quickly decays to $MAB'$ and $C$, and finally to $M$ and $AB$.  
For clarity, we have not shown $MABC$ in Fig. \ref{rate}.  
We distinguish $MAB'$ from $MAB$ (in the upper branch)
because in our model branch migration is assumed to occur more readily in the former, so it has a faster
decay rate than $MAB$.  The observed fluorescence 
intensity is proportional to the sum of all non-free M, $[MA]+[MAB]+[MAB']$.  

As implied in Fig. \ref{rate}, the fast and slow reactions compete 
for $B$.  Our experiment shows that the competition is strongly tilted towards the lower branch
as soon as the catalyst concentration $c_0$ is increased from zero (steep drop in 
$\tau$ in Fig. \ref{tau}).  The saturation of $\tau$ at large $c_0$ also implies that 
the slow process becomes insignificant for $c_0>$ 0.5 $\mu$M.  Hence the 
curves at 1.5 $\mu$M are determined by $k_0$, $k_3$ and $k_4$.  It turns out
that $k_4$ is the most important parameter determining the observed decay 
in the high-$c_0$ regime.  The ``bottleneck'' at the node $MAB'$ resulting from
the relatively small value of $k_4$ compared with $k_3$ makes the decay rate 
rather insensitive to the latter (the lines in Fig. \ref{rate} are drawn with widths roughly
proportional to the rate constants).  By fitting the curves in the 2 
extreme limits, $c_0 = 0$ and $c_0$ = 1.5 $\mu$M (Appendix), we find the 
values of $k_1$, $k_2$ and $k_4$ to be 
\begin{eqnarray}
k_1 &=& (0.21\pm 0.04) \quad \mathrm{\mu M^{-1}s^{-1}}, \label{k1}\\
k_2 &=& (8.9\pm 1.0)\times 10^{-3}\;\mathrm{s^{-1}},\label{k2}\\
k_4 &=& (3.4\pm 0.1)\times 10^{-2}\;\mathrm{s^{-1}}. \label{k4}
\end{eqnarray}
[We remark that the second-order rate constants $k_0$, $k_1$ and $k_3$ 
(reactions on the left half of Fig. \ref{rate}) carry an extra
factor of concentration compared with the first-order constants $k_2$ and $k_4$
(right half).  Since the concentrations used here are nominally $\mu$M, we have
expressed the former in units $\mathrm{\mu M}^{-1}$ s$^{-1}$, and the latter 
in s$^{-1}$.   In these units, the numbers quoted provide a more meaningful 
comparison of the reaction rates on the left half with those on the right.]

While the remaining rate constants $k_0$ and $k_3$ only weakly affect the
decay rates of the large-$c_0$ curves, they obviously influence the strong competition
for $B$ when $c_0$ is small.  Hence the observed curve of $\tau$ vs. $c_0$ in Fig. \ref{tau}
may be used to estimate 
\begin{eqnarray}
k_0 &=& (2 \pm 1) \quad\quad\quad\mathrm{\mu M^{-1}s^{-1}}, \label{k0}\\
k_3 &=& (2.5\pm 0.5) \quad\quad \mathrm{\mu M^{-1}s^{-1}}.\label{k3}
\end{eqnarray}
Although the uncertainties in $k_0$ and $k_3$ are relatively large, 
our fits exclude values of $k_0 < 1\;\mathrm{\mu M^{-1}s^{-1}}$ and $k_3\;<$ 2 
$\mathrm{\mu M^{-1}s^{-1}}$.  The large value of $k_3$ relative to $k_4$
leads to the bottleneck at the node $MAB'$ mentioned above.

The ratio of the reaction rate constants $k_4/k_2\sim 3.8$ 
implies that the restoring reaction Eq. \ref{MAB} is significantly improved with the
addition of $C$.  The large value of $k_3$ ($\sim 10 k_1$) implies that $C$ is highly efficient
in catalyzing the formation of the complex $MAB'$ compared
with the formation of $MAB$ in the catalyst-free solution.  This 
presumably reflects the inhibition of the folding of $B$.

\section{Summary}
In summary, we have designed a DNA catalyst strand that significantly accelerates 
the restoration reaction time in the device proposed by Li and Tan.  
Recently, Dittmer, Reuter and Simmel~\cite{Dittmer} reported an interesting 
application of a DNA aptamer similar to this device.  They successfully used
the machine to precisely control the concentration of thrombin protein in solution between
a depleted and an enriched state.   Our catalyst may be useful in their 
application.  A model incorporating the reaction rate constants of the intermediate product helps us to 
understand the rate-limiting features in the operation of this single-molecule device.  
As noted in Ref. ~\cite{Turberfield}, the catalyst itself is also a DNA machine that obtains energy by 
catalyzing the restoration reaction.  Such machines can run until the supply of 
unreacted fuel is exhausted.

\section*{Appendix}
We discuss in more detail our analysis, and the optimization procedure.  
First, we justify the neglect in Eqs. \ref{BC} and \ref{MA} of the off reaction-rate constants $k_{-}$ 
(defined in Eq. \ref{SS}).  We have 
\be
K \equiv \frac{k_{+}}{k_{-}}\times 1\mathrm{M} = \mathrm{e}^{-\Delta G/RT},
\label{kk}
\ee
where $\Delta G$ is the 
reaction Gibbs free energy, $R$ the molar gas constant
and $T$ the absolute temperature.
The Gibbs free energies 
are estimated from the data of Peyret and SantaLucia~\cite{Peyret}.  
In a solution at 20$^{\mathrm{o}}$C with 1 M monovalent cation, the Gibbs 
free energies are $\Delta G_{MA}$ = -27.55 kcal/mol (for hybridization of $M$ and $A$), 
$\Delta G_{AB'}$ = -16.13 kcal/mol (of $A$ and $B$) and $\Delta G_{BC}$ 
= -13.40 kcal/mol ($B$ and $C$).  In our experiment the nominal values of $[S]_0$ 
are 1 $\mu$M, which implies that $k_{+}[S]_0/k_{-}\gg 1$ in each reaction 
(the smallest value -- $k_{+}[S]_0/k_{-} =$ 9,900 --
is obtained in the reaction Eq. \ref{BC}).  Hence, in comparison with the forward
rate constants, we may ignore $k_{-}$ in Eqs. \ref{BC} and \ref{MA}~\cite{Yurke4}.
In the reactions Eqs. \ref{MAB}-\ref{M}, the reverse reactions, which involve the dehybridization of 
the stable complexes $MAB'$ and $AB$, have a negligibly small probability as well.

In terms of the the forward rate constants, the time derivatives of the populations are
\begin{eqnarray}
\frac{d[BC]}{dt} &=& k_0[B][C] \label{d0}\\
\frac{d[MAB]}{dt} &=& k_1[MA][B] - k_2[MAB] \label{d1} \\
\frac{d[MAB']}{dt} &=& k_3[MA][BC] - k_4[MAB'] \label{d2} \\
\frac{d[M]}{dt} &=& k_2[MAB] + k_4[MAB']. \label{d3}
\end{eqnarray}

In the experiment, the total amount of $M$, $A$ and $B$ add up to 1.0 $\mu$M, while
the amount of $C$ is $c_0$.  Hence by mass conservation, we have (with the assumption
$[A]$ = 0)
\be
[M] + [MA] + [MAB] + [MAB'] = 1.0 \;\mu\mathrm{M} \label{M1}
\ee
\be
[MA] + [MAB] + [MAB'] + [AB] = 1.0 \;\mu\mathrm{M} \label{M2}
\ee
\be
[B] + [BC] + [MAB] + [MAB'] + [AB] = 1.0 \;\mu\mathrm{M} \label{M3}
\ee
\be
[C] + [BC] = c_0.\label{M4}
\ee
Equations \ref{M1} and \ref{M2} imply $[M] = [AB]$.

For a given $c_0$, we have as input 
the measured fluorescence intensity curve (normalized to 1 at $t =0$
before the addition of $B$)
\be 
f(t) = \frac{I(t) - I_{\infty}}{I(0)-I_{\infty}}
\label{ft}
\ee
(with the caveat that data is missing in the blind interval $\Delta t$).  
For a set of starting values of $k_i$'s we integrate numerically
the rate equations to determine the curve of $1-[M](t)$ (hereafter, 1 means 1.0 $\mu$M).  
A least-squares fit is used to fit the curve of $1-[M]$ to $f(t)$ as follows.

We note that the rate equations 
Eqs. \ref{d1}-\ref{d3} simplify in the 2 limits of zero concentration and high concentration of $C$.  
In the first case, we have $[BC] = [MAB'] = 0$, which gives (from Eqs. \ref{M1} and \ref{M3})
\begin{equation}
[B] = [MA] = 1-[M]-[MAB]\quad(c_0 = 0).
\end{equation}
Equations \ref{d1}-\ref{d3} reduce to
\begin{eqnarray}
\frac{d[MAB]}{dt} &=& k_1(1-[M]-[MAB])^2 - k_2 [MAB] \label{dx}\\
\frac{d[M]}{dt} &=& k_2 [MAB].  \label{dy}
\end{eqnarray}

In the curve-fitting procedure, we assume seed values for $k_1$ and $k_2$, and integrate 
Eqs. \ref{dx} and \ref{dy} numerically to obtain the curve of $y(t) = 1-[M](t)$, 
which is then compared with the measured $f(t_{obs})$.
Here, $t_{obs}  = t - \Delta t$ where $\Delta t$ is the slight 
shift in the origin of the time axis caused by the 
mixing step, as explained in Sec. \ref{expt}.  
The values of $k_1$, $k_2$ and $\Delta t$ are successively refined
until convergence is attained.  The convergence 
is faster if we regard $y$ and $f$ as the independent variables,
and perform the least-squares fit on the inverted curves $t(y)$ 
and $t_{obs}(f)+\Delta t$. For the 3 samples with $c_0 = 0$, the 
optimization process yields $k_1 = (0.21\pm 0.4) \; \mu\mathrm{M^{-1}s^{-1}}$,
$k_2 = (8.9\pm 1.0)\times 10^{-3}\;\mathrm{s^{-1}}$ and 
$\Delta t_0 = 56$ s.  

By fitting to the curves with $c_0$ = 1.5 $\mu$M, we 
find $k_4 = (3.4\pm 0.1)\times 10^{-2}\;\mathrm{s^{-1}}$, 
and $\Delta t_1$ = 21 s.
(As noted, these large-$c_0$ curves are highly sensitive
to variations in $k_4$, but only weakly sensitive to variations in $k_0$ and $k_3$.)
With the values of $k_1$, $k_2$ and $k_4$ determined, we 
can calculate the effective time constant $\tau$ vs. $c_0$ using assumed values of 
$k_0$ and $k_3$.  Finally, variation of these values allows us to optimize the fit of 
$\tau$ vs. $c_0$ to the measurements in Fig. \ref{tau}.  The best values are
reported in Eqs. \ref{k0}-\ref{k3}.

We thank Zong Lin and Hays S. Rye for assistance with the fluorescence 
measurements.  We also thank R. H. Austin for fruitful discussions.  This research
is supported by Princeton University.

$^\dagger$ \emph{Present address of Y. Zhang: Biosciences Division, Argonne National Lab.,
Argonne, IL 60439}

\bfig
\incl[width=9cm]{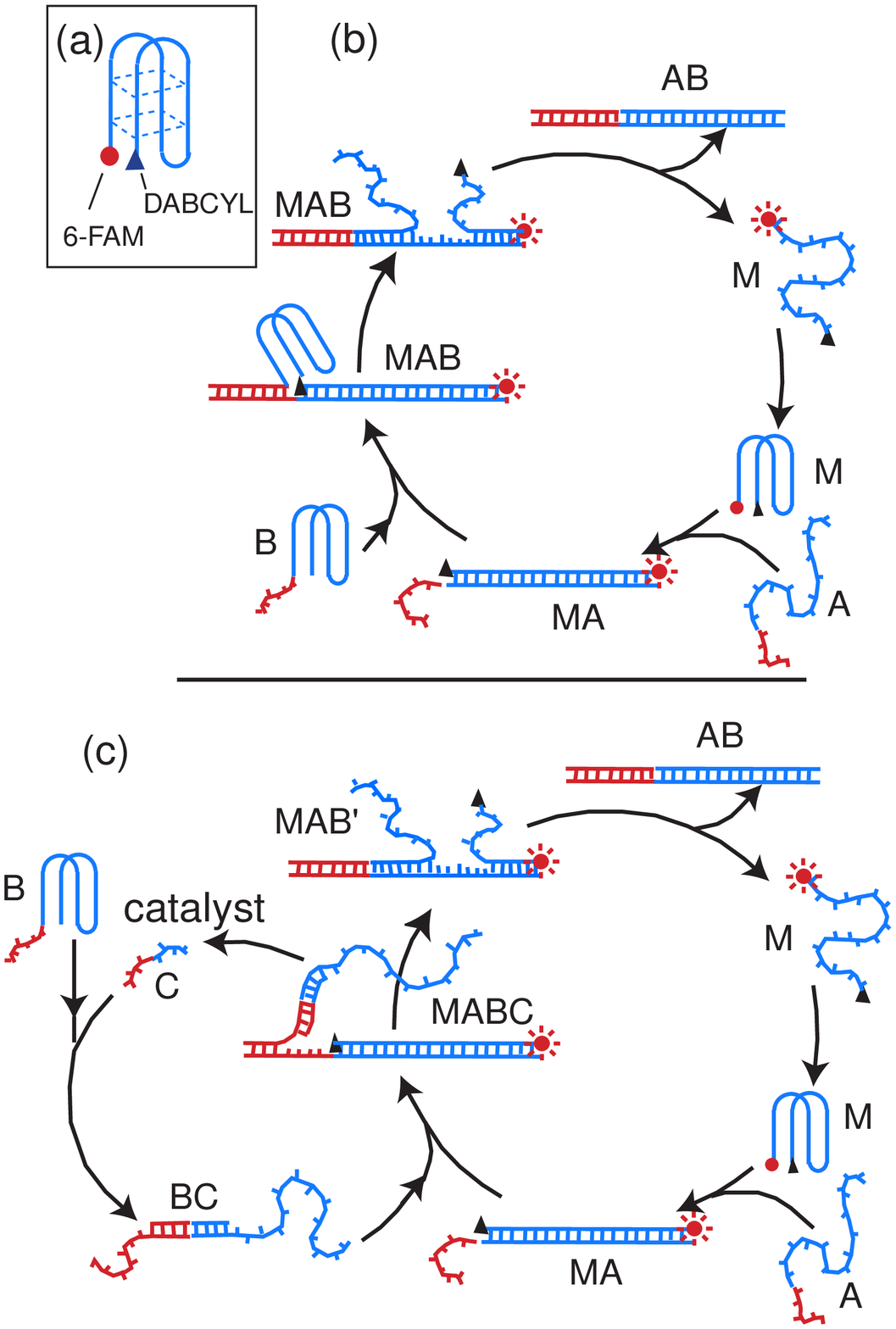}
\caption{\label{strands} Operation of the single-molecule DNA device. 
(a) The motor strand $M$ in solution.  $M$ is labeled at end 5' with 
the fluorophore 6-FAM (5'-fluorescein phosphoramidite) 
and at end 3' with the quencher DABCYL [4-(4'-dimethylaminophenylazo) 
benzoic acid]. When folded into the chair structure, close proximity of the two ends 
quenches fluorescence.  Dashed lines indicate the two G-quadruplexes in folded configuration.
(b)  Reaction cycle of device introduced by 
Li and Tan~\cite{Tan}.   In the `straightened' configuration 
of $MA$, the fluorescence intensity is high.  The intensity decreases as 
folding is gradually restored with introduction of the restoration 
fuel $B$.  Undesired folding of $B$ slows the restoration reaction. 
(c) Reaction cycle with added catalyst. The catalyst $C$ is designed to 
bind to $B$ to inhibit folding.  It also helps to initiate the branch 
migration and accelerate the restoration reaction. For substoichiometric 
$c_0$ both (b) and (c) occur when $B$ is added.  For 
stoichiometric and above-stoichiometric $c_0$ (c) dominates the 
restoration reaction.
}
\efig
\bfig
\incl[width=9cm]{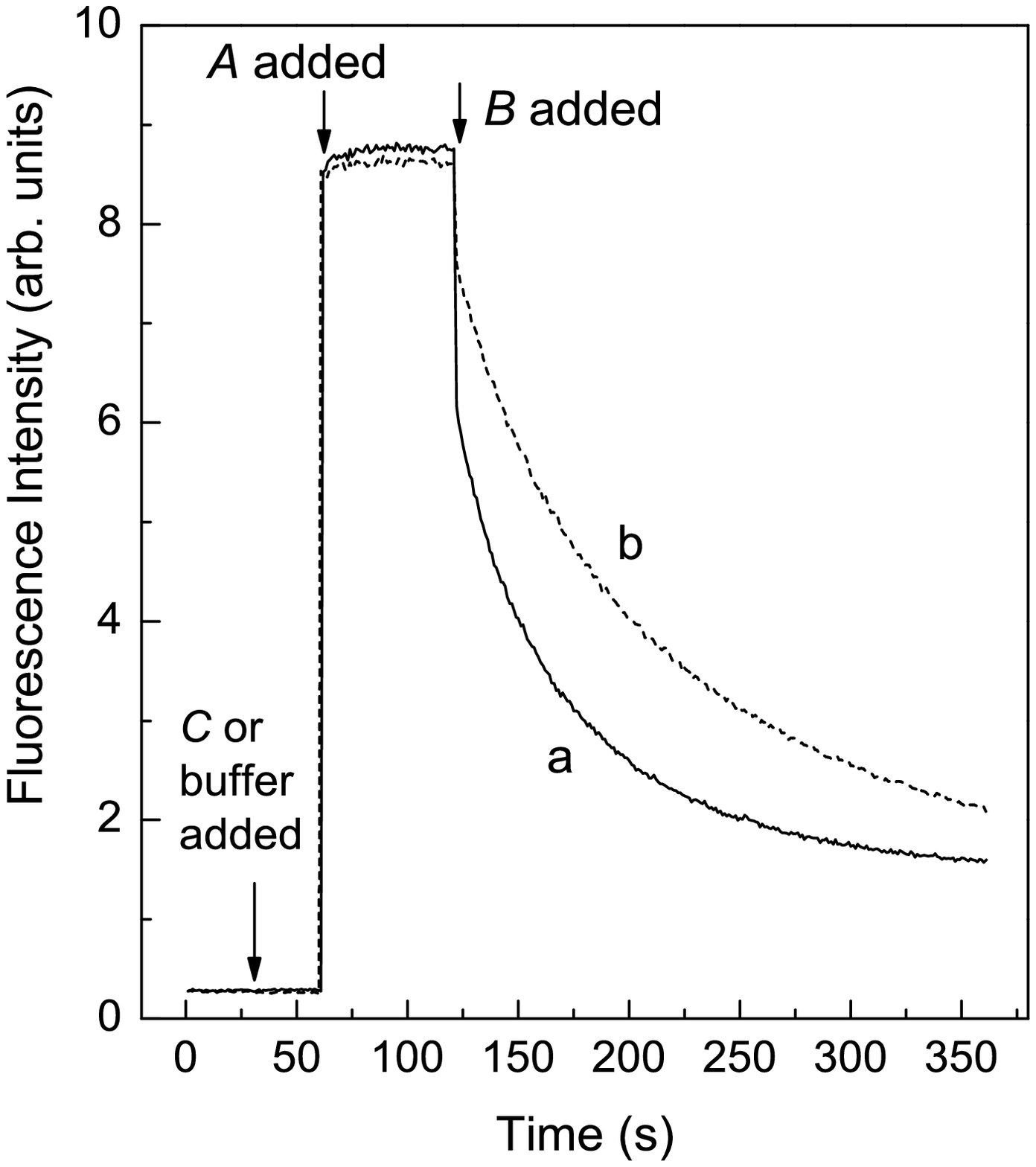}
\caption{\label{decay}
A reaction cycle of the single DNA molecule motors with (Sample $a$) and without 
(Sample $b$) the catalyst.  High fluorescence signal corresponds to having
the ends of $M$ far apart while low fluorescence means that the ends are in close
proximity ($M$ folded).  In each curve, the sample is 100 $\mu$l of the 
motor strand $M$ at the concentration 1 $\mu$M in KSPSC buffer at $t = 0$.  
At time $t$ = 30 s, 1 $\mu$l of 100 $\mu$M $C$ stock solution was added 
to Sample $a$, while 1 $\mu$l of SPSC buffer was added to the control 
sample $b$ (i.e. the catalyst concentration is $\sim 1.0\;\mu$M in $a$ 
and 0 in $b$).  Note that the catalyst accelerates 
the restoration reaction without affecting the straightening reaction.  
Right before $A$ or $B$ is added, the shutter of the spectrofluorometer was closed 
(for $\sim$10 s) and the solution rapidly pipetted.  This 
accounts for the discontinuity at $t $ = 60 and 120 s (see text). 
}
\efig
\bfig  
\incl[width=9cm]{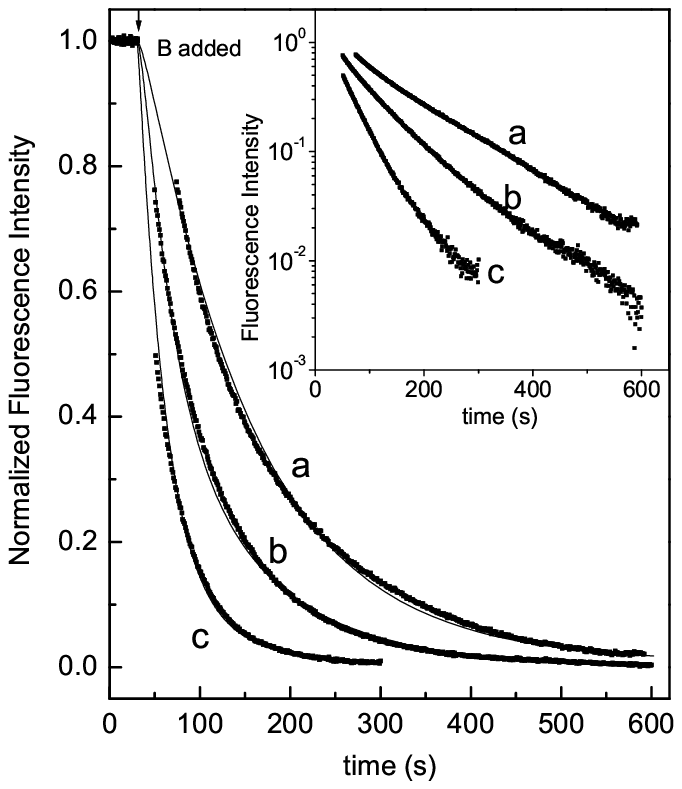}
\caption{\label{modelfit}
Time-dependence of 6-FAM fluorescence during the restoration reaction. The background quenched intensity from 
folded $M$ has been subtracted, so the fluorescence 
intensity is proportional to the sum of all non-free $M$, $[MA]+[MAB]+[MAB']$.  
The intensity before addition of $B$ has been normalized to 1.  The solid lines are fits to the 
Eqs. \ref{BC}-\ref{M} (see Appendix).  The calculated curves with optimized reaction 
rate constants $k_1\cdots k_4$ interpolate within the `blind' intervals $\Delta t$.  The catalyst 
concentration in Samples $a$, $b$ and $c$ are 0, 0.1 and 1.5 $\mu$M $C$, 
respectively.  In the inset, the same data are displayed in semi-log scale.  
The initial straight portion is used to determine an effective time constant $\tau$.
}
\efig
\bfig  
\incl[width=9cm]{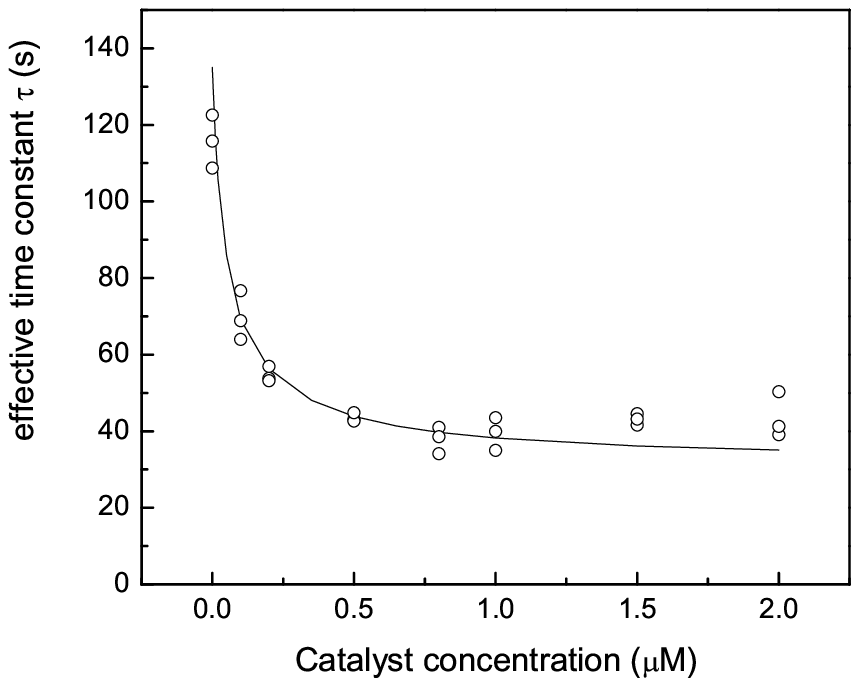}
\caption{\label{tau}
Restoration reaction time constant as a function of catalyst concentration 
([M] = [A] = [B] = 1.0 $\mu$M).  As noted in the text, values of $\tau$ shown here are inferred 
directly from semi-log plots of the raw data, and are independent of 
fits to the model.  The best fit to $\tau$ vs. $c_0$ (solid curve) is used to determine the reaction
rate constant $k_3$ (Appendix).
}
\efig 
\bfig 
\incl[width=9cm]{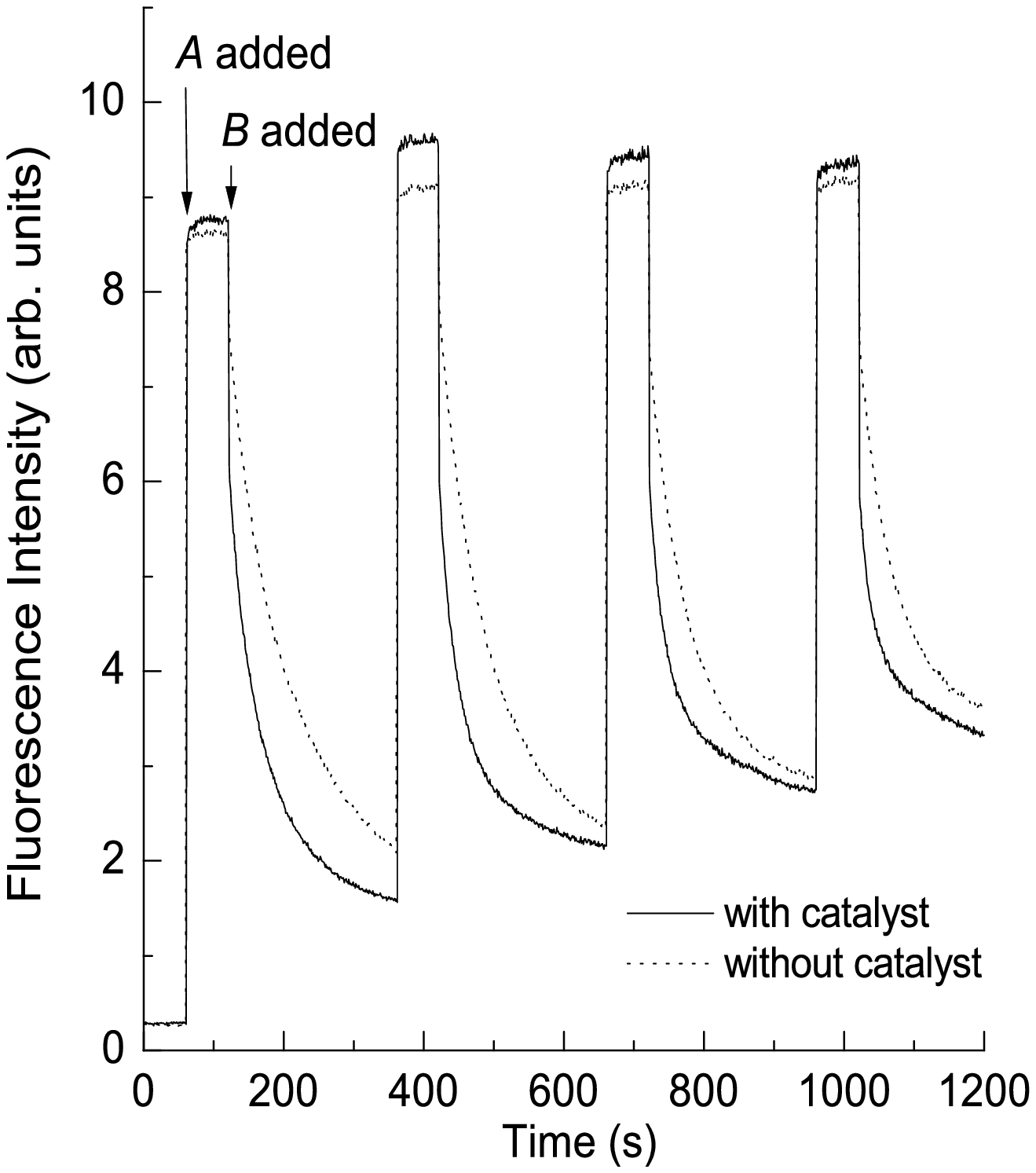}
\caption{\label{repeatcycles}
Cycling the DNA device with 1.0 $\mu$M of catalyst $C$ (solid curve) and without catalyst (dotted curve).  
In each cycle, high (low) fluorescence signal indicates straightening (folding) of the motor strands $M$.
The catalyst remains efficient for several cycles.
}
\efig
\bfig 
\incl[width=7cm]{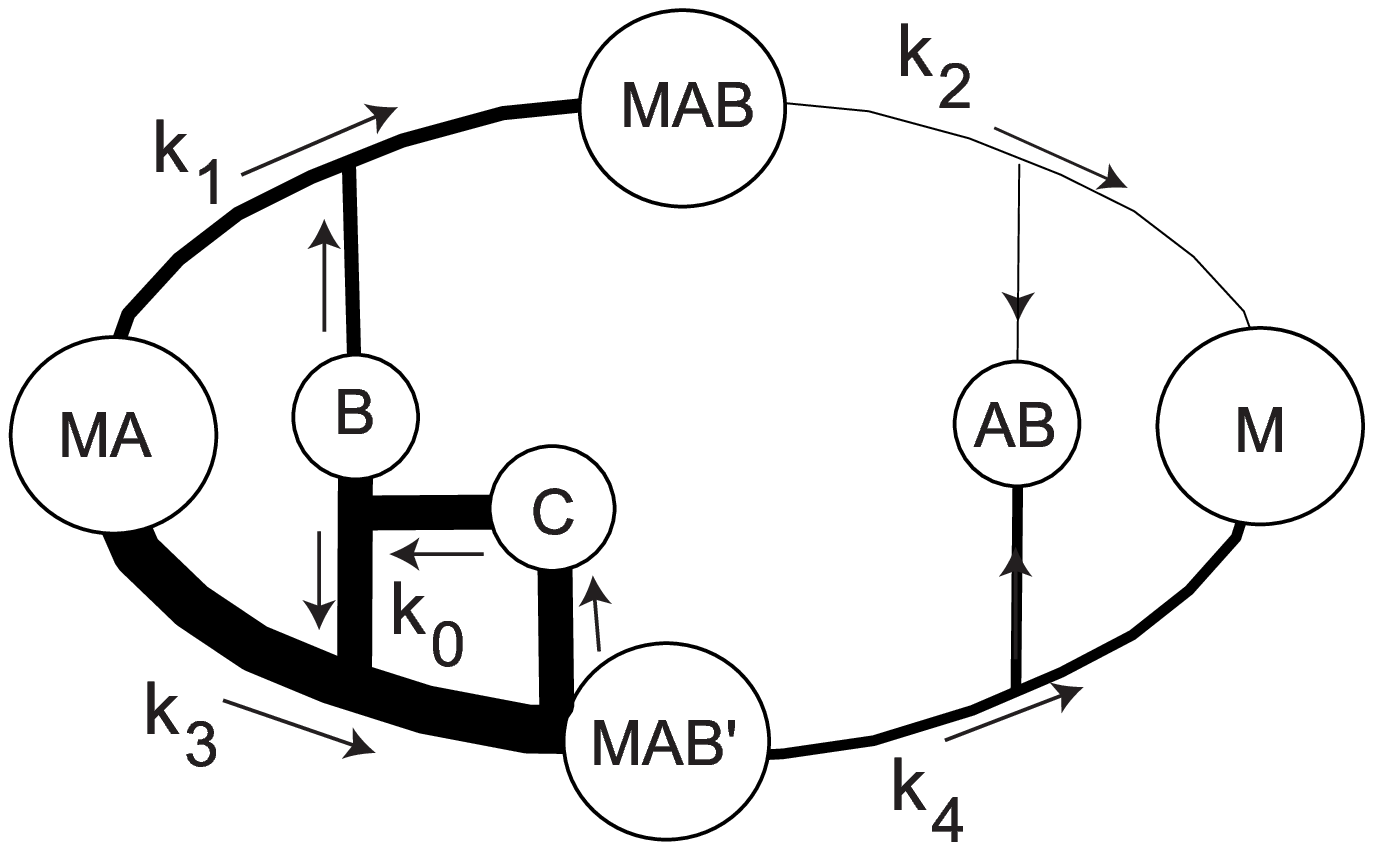}
\caption{\label{rate}
Schematic drawing of the reaction paths in Eqs. \ref{BC}-\ref{M}.  The duplex $MA$, fuel strand $B$
and catalyst $C$ are `sources' on the left while the motor strand $M$ and waste product $AB$ are sinks on the right.
In the absence of $C$, the reaction proceeds along the upper branch (with rate constants $k_1$ and $k_2$).
Introducing $C$ turns on the process on the lower branch.  For small $c_0 (<0.5\; \mu$M), the 2 
reactions strongly compete for $B$, whereas for $c_0> 0.5\;\mu$M, the lower branch dominates.  
The thickness of the connecting lines nominally reflects the rate constants.  In both branches, the small values 
of the rate constants $k_2$ and $k_4$ lead to bottle-necks in the intermediate 
products $MAB$ and $MAB'$. 
}
\efig

\end{document}